\documentclass[12pt]{article}
\usepackage[latin1]{inputenc}
\usepackage{amssymb}
\usepackage{graphicx}
\usepackage{float}
\usepackage{hyperref}
\newcommand{\text}{\rm}

\newcommand{\bb}{\begin{equation}}
\newcommand{\ee}{\end{equation}}
\newcommand{\bega}{\begin{eqnarray}}
\newcommand{\ega}{\end{eqnarray}}
\newcommand{\begae}{\begin{eqnarray*}}
\newcommand{\egae}{\end{eqnarray*}}

\newcommand{\h}{\hspace*{4ex}}

\newcommand{\cent}{\centerline}
\newcommand{\vs}{\vspace*}

\begin{document}

\baselineskip 0.5cm

\begin{center}

{\large {\bf Holographic metasurfaces simulations applied to realization of non-diffracting waves in the microwave regime } }

\end{center}

\vs{0.2 cm}

\cent{Santiago R. C. Fernandez$^{\: 1}$, and Marcos R. R. Gesualdi$^{\: 1}$}

\vs{0.2 cm}

\centerline{{\em $^{\: 1}$ Universidade Federal do ABC, Av. dos Estados 5001, CEP 09210-580, Santo Andr\'e, SP, Brazil.}}

\vs{0.5 cm}

{\bf Abstract  \ --} \ In this work, we present the computational realization of holographic metasurfaces to generation of the non-diffracting waves. These holographic metasurfaces (HMS) are simulated by modeling a periodic lattice of metallic patches on dielectric substrates with sub-wavelength dimensions, where each one of those unit cells alter the phase of the incoming wave. We use the surface impedance (Z) to control the phase of the electromagnetic wave through the metasurface in each unit cell. The sub-wavelength dimensions guarantees that the effective medium theory is fulfilled. The metasurfaces are designed by the holographic technique and the computer-generated holograms (CGHs) of non-diffracting waves are generated and reproduced using such HMS in the microwave regime. The results is according to the theoretically predicted by non-diffracting wave theory. These results are important given the possibilities of applications of these types of electromagnetic waves in several areas of telecommunications and bioengineering. \\


\vs{0.5 cm}

\h {\em\bf 1. Holographic Metasurfaces and Surface Impedance} --- The artificially structured materials such as photonic crystals and metamaterials have attracted great interest for their remarkable properties to control and manipulate light and electromagnetic waves \cite{ref:1,ref:2}. Metamaterials are artificial materials composed by a periodic array of sub-wavelength unit cells, they have been very explored in a wide range of applications due to fact of their properties depends on the geometry and materials of their unit cells \cite{ref:3,ref:4,ref:5,ref:6,ref:6a}. The requirement of sub-wavelength dimensions is important for approximation of effective media be fulfilled, it means, the incoming wave can't distinguish the discontinuities of the metamaterial and therefore, it can be considered as an homogeneous media and characterized by effective parameters of permittivity and permeability. The novelty in those metamaterials is the capacity of having permittivity and permeability simultaneously negatives in a same interval of frequencies. Some important applications of metamaterials are the obtaining of negative refractive index with resonant character \cite{ref:5,ref:7}, superlensing \cite{ref:7a,ref:7b}, the phenomena of negative refraction \cite{ref:8,ref:9,ref:10} and the possibility of cloaking light around certain physical spaces \cite{ref:11,ref:12}. \\

\h The applications of three-dimensional metamateriais can be also applied to their two-dimensional versions: the metafilms or metasurfaces. Artificial surfaces composed by a periodic array of sub-wavelength unit cells or resonators \cite{ref:13,ref:14}. Metasurfaces offer great advantages over their analogous three-dimensional metamaterials, as for example, less losses, more comfortability, occupying less physical space and more connectivity to conventional equipments in laboratories. \\

\h Metasurfaces are fundamental devices to control or modify wavefront, phase or polarization state. The resonators introduce abrupt changes of phase in the interface due to the discontinuities on the surface. The result is a generalization of the laws of the reflection and refraction, being possible the control of a refracted wave by modulating the gradient of phase imposed by the resonators \cite{ref:15,ref:16}. Thus, metasurfaces provide us to shape the wavefront in shapes designed at will only by building and by ordering the suitable resonators \cite{ref:17}. In the designing such metasurfaces is very important to have a complete control of the phase of the wave scattered by resonators (from 0 to $2\pi$), for achieving this complete interval, much type of resonators have been proposed depending on the required functionality. As for example, arrays of plasmonic nanoantennas with variation of angles and orientation for achieving tunable amplitude and polarization state of the refracted wave \cite{ref:18,ref:19,ref:20}. \\

\h On other hand, the holography was developed as a method for reconstruct wave fronts and producing three-dimensional images. Through process of registration, the information of phase and amplitude of a wave scattered for the surface of an object (object wave) is stored in determinate photosensitive materials by interference with the reference wave. Thus, the hologram captures the interference pattern between those two optical waves. The reconstruction occurs when the hologram is illuminated by reference wave, the diffraction pattern reproduces the wavefront from the original object \cite{ref:23,ref:24,ref:25}. \\

\h In this work, we use the surface impedance ($Z$) to control the phase of a wave through the metasurface by controlling each unit cell, such metasurface is called holographic metasurface (HMS) \cite{ref:21,ref:22}. That name comes from holographic principle, where the interference happens between the surface wave $\psi_{\textrm{surf}}$, the incoming wave passing through the surface (reference wave), and the radiation wave $\psi_{\textrm{rad}}$, the transmitted wave from the surface (object wave). For reconstructed the radiation wave, we use the surface wave to excite the interference pattern: $(\psi_{\textrm{surf}}^{*}\psi_{\textrm{rad}})\psi_{\textrm{surf}}=\psi_{\textrm{rad}}|\psi_{\textrm{surf}}|^{2}$. Thus, to realize the radiation pattern, we need a distribution for a surface impedance on the metasurface, the theoretical equation for the surface impedance is given by interference of $\psi_{\textrm{surf}}$ and $\psi_{\textrm{rad}}$ \cite{ref:22} 

\begin{equation}\label{e0}
Z=i\left[X+M{Re}(\psi_{\textrm{rad}}\psi_{\textrm{surf}}^{*})\right]
\end{equation}

where $X$ and $M$ are modulation values. \\

\h The impedance surface ($Z$) is defined as the ratio between the component of electric field parallel to a current along the surface and the current per unit length of surface: $\mathbf{E}_{t}=Z(\mathbf{x}_{t})\mathbf{J}$. For the case of a metasurface, we should average that equation over the unit cell, the result for the TM modes (transverse magnetic waves) is given by \cite{ref:22}: 

\begin{equation}\label{e1}
Z=iZ_{0}\biggl(\frac{k_{z}}{k}\biggl)
\end{equation}

where $Z_{0}$ is the vacuum impedance, $k$ the wave vector and $k_{z}$ the decay constant, considering the surface wave as $Ae^{-i(\mathbf{k}_{t}\cdot\mathbf{x}_{t})-k_{z}z}\ e^{i\omega t}$. We can obtain another result for the surface impedance (\ref{e1}) by finding a value for the transverse wave vector, $k_{t}$. Using the software CST Microwave Studio, we can simulate a unit cell with lattice parameter $d$ and to find the phase ($\phi$) through such unit cell for the frequency $\omega$, according to $\phi=k_{t}d$: 

\begin{equation}\label{e2}
\left(\frac{k_{z}}{k}\right)^{2} = \left(\frac{k_{t}}{k}\right)^{2}-1 = \left(\frac{\phi/d}{\omega/c}\right)^{2}-1
\end{equation}

being $c$ the light speed at vacuum. Then, we can calculate the expression that relates the surface impedance and the phase through a unit cell: 

\begin{equation}\label{e3}
Z = Z_{0}\sqrt{1-\phi^{2}c^{2}/\omega^{2}d^{2}}
\end{equation}

\begin{figure}[H]
\centering
\includegraphics[width=9cm]{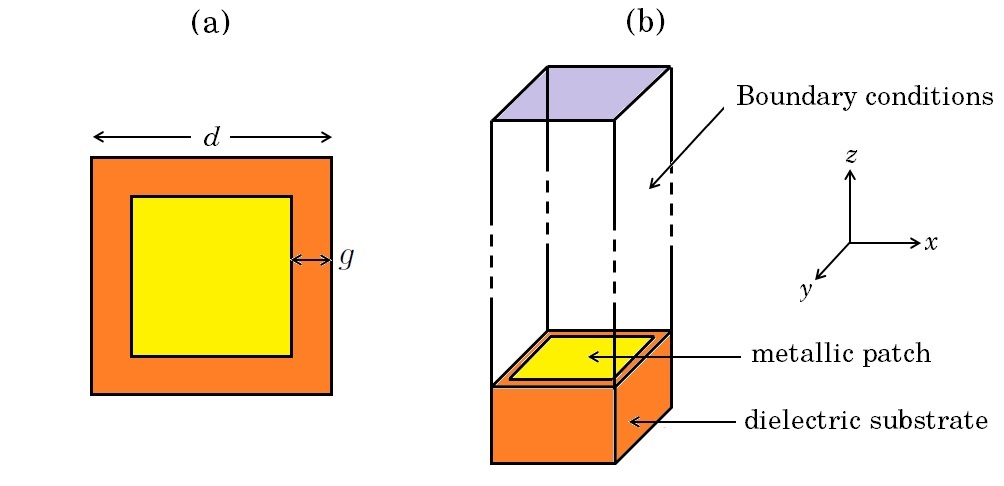}
\caption{\textbf{(a)} Unit cell of the metasurface. \textbf{(b)} Boundary conditions in the unit cell designed in CST.}
\label{f1}
\end{figure}

\h In this way, we design the holographic metasurface (HMS) consisting on a set of unit cells, each one of them formed by a metallic patch on a dielectric substrate, both elements have square shape being the side of metallic patch always less than the side of substrate, thus, a gap ($g$) is formed in each unit cell (see Figure \ref{f1}) \cite{ref:40}. We can choose a determinate numbers of gaps for making the simulations, usually nine or ten values equally spaced are chosen, from a $g_{\textrm{min}}$ till $g_{\textrm{max}}$. For every value of gap, we design in CST Microwave Studio software the corresponding unit cell, applying the function Eingenmode solver to obtain the dispersion curve, i.e. the variation of frequency with the phase. We repeat this procedure for each value of gap previously chosen, obtaining a graph for each one of them. When all those graphs are superimposed we choose the operating frequency ($f$), it is defined so that we be able to find a unique value of phase for every value of gap at such frequency. For waves in the microwave regime, the wavelengths varies from 1 mm to 1 m, therefore the dimensions of our holographic metasurfaces must be in order of millimeters (mm) regarding $d\lesssim\lambda/10$ for is in according with the effective media theory. \\

\h We can to obtain for a frequency $f$ the value of $Z$ corresponding to a value of phase $\phi$, according to the equation \ref{e3}, and at same time, corresponding to a value of gap $g$. This result is very useful because it allows us to control the surface impedance $Z$ just varying the values of gaps. Thus, we can obtain an interval for surface impedance $[Z_{\textrm{min}}, Z_{\textrm{max}}]$ starting from the chosen values of gaps $[g_{\textrm{min}}, g_{\textrm{max}}]$. For having continuous values of $Z$ for any value of $g$, we can make an adjustment of the discrete points $Z$ vs $g$ by interpolations and, finally to get the relation $g = g(Z)$. \\

\h Then, if we have a map of values of surface impedance $Z$ for every unit cell of the holographic metasurface, we can find the corresponding value of gap and design the whole metasurface. Those values of $Z$ are given by the expression \ref{e0}, and therefore, they will depend on type of wave that we want obtain through the metasurface. In this work, we want to obtain the holographic metasurfaces (HMS) of computer-generated holograms (CGH) of object waves previously calculated. Thus, the interference pattern in the expression \ref{e0} would be in the own CGH, i.e. the information of phase is given in each pixel of CGH. Every pixel of the computer-generated hologram has a gray level between 0 (black) and 255 (white), we associate each one of these values to a value of phase between 0 and $2\pi$ for obtaining a matrix of phase ($\Phi$) for the selected CGH. Thus, according to the expression \ref{e0}, we have: 

\begin{equation}\label{e4}Z = i[X+M\Phi]\end{equation}

and now, $X$ and $M$ would be adjustment values for making $Z$ to be inside the interval previously calculated for surface impedance $[Z_{\textrm{min}}, Z_{\textrm{max}}]$, making the calculations, we found: $X=Z_{\textrm{min}}$ and $M$ is the maximum integer value satisfying: $M\leq(Z_{\textrm{max}}-Z_{\textrm{min}})/2\pi$. \\

\h Thus, we have a matrix of surface impedance for every pixel of the CGH of the non-diffracting wave and and the relation $g = g(Z)$. Therefore, we have a unique value of gap for each pixel of the CGH and thus, we can build an array of such unit cells forming the holographic metasurface working in the microwave regime. Following are shown the results obtained for some CGH of known wavefronts. Particularly, in this work this wavefronts is the non-diffracting waves.

\h {\em 2. Non-diffracting waves} --- Non-diffracting waves are beams and pulses that keep their intensity spatial shape during propagation \cite{ref:26,ref:28,ref:29,ref:30,ref:31,ref:32,ref:33,ref:34,ref:35,ref:36,ref:37,ref:39}. Pure non-diffracting waves include Bessel beams, Mathieus beams and Parabolic beams; as well as the superposition of these waves can produce very special diffraction-resistant beams, such as the Frozen Waves. These waves could be applied in many fields in photonics. \\

\h In optics, the experimental generation of non-diffracting beams using conventional diffractive optical components presents several difficulties, as co-propagating beam superposition for instance, and in some cases, is not feasible. Thus, a type of holography very important and relevant lately, is the computer-generated holography (CGH). In this case, the hologram is created from computational numerical methods. The physical process that allows the reconstruction of the image in far-field is expressed by the theory of diffraction of Fresnel-Kirchhoff \cite{ref:24}. The computational holography technique with the use of numerical holograms and spatial light modulators, has efficiently reproduced the beams cited above. In this case, the construction of the non-diffracting beam hologram is done numerically by a computer generated hologram (CGH) and its reconstruction is performed optically with its implementation in a spatial light modulator (SLM). \\

\h In this work, we will focus on computer-generated holograms of phase \cite{ref:25} of some types of non-diffracting waves. The non-diffracting waves are solutions to the (linear) wave equation which travel well confined or {\em localized\/}, in a single direction without to experiment effects of dispersion caused by diffraction. The types of non-diffracting waves studied are Airy beams \cite{ref:26,ref:35}, Bessel beams \cite{ref:28,ref:29} and Frozen waves (FW) \cite{ref:30,ref:31,ref:32,ref:33,ref:34,ref:35,ref:36,ref:37,ref:39}. \\

\h {\em 3. Simulations and Results} --- We built two sets of holographic metasurfaces working in different frequencies in the microwave regime by simulations using CST Microwave Studio software \cite{ref:40}. 

\begin{figure}[H]
\centering
\includegraphics[width=13cm]{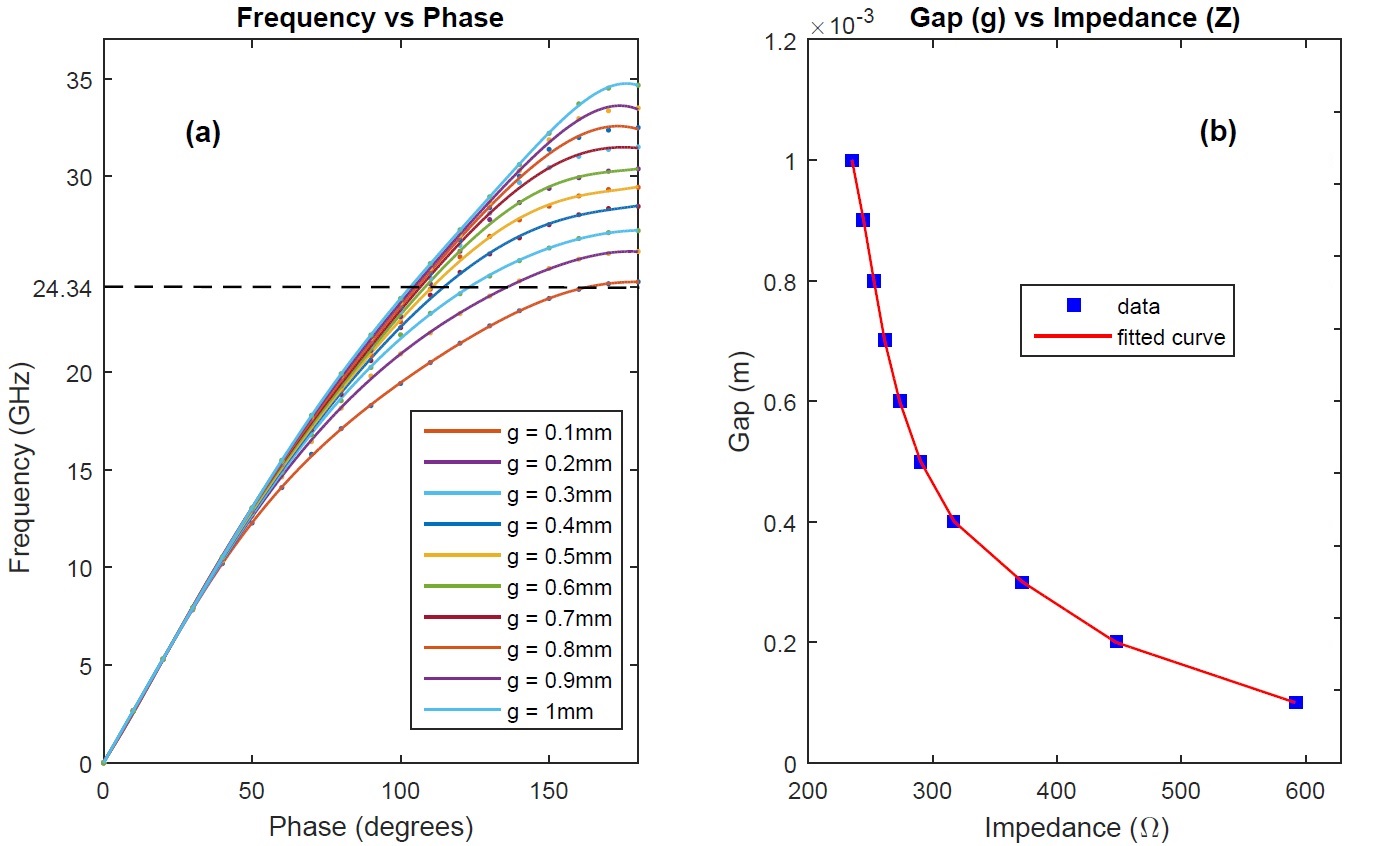}
\caption{(a) Variations of frequency with phase for each v
alue of gap ($g$), the operating frequency was defined to 24.34 GHz. (b) Variation of values of gaps according to surface impedance and its corresponding adjustment curve.}
\label{f2}
\end{figure}

\h The first case, at operating frequency of 24.34 GHz, the metal used was copper (Cu) on a substrate Rogers RT5880 with permittivity $\epsilon=2.2$ and thickness $t=1.57$ mm. The lattice parameter was set at $d=3$ mm, and gaps were set from $g=1$ mm till $g=2.8$ mm with an interval of 0.2 mm. The interval of frequencies was defined from 0 till 100 GHz. In the four sides of the unit cell (see figure \ref{f1}) were set periodic boundary conditions and Perfect Electrical Conductor (PEC) for the propagation direction. The dispersion curves for each value of gap and the variations of gaps versus surface impedance are shown in the Figure \ref{f2}. After making the adjustment of the curve of $g$ vs $Z$, we obtained the interval for surface impedance: $[Z_{\textrm{min}}, Z_{\textrm{max}}] = [235.05\ \Omega, 591.39\ \Omega]$, and we also found the following modulating values: $X=235.05\ \Omega$, $M=56$. \\

\begin{figure}[H]
\centering
\includegraphics[width=5cm]{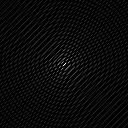}
\caption{Image of the computer-generated hologram of a Bessel beam with resolution of 128x128 pixels at wavelength $\lambda=12.33$ mm.}
\label{f4}
\end{figure}

\begin{figure}[H]
\centering
\includegraphics[width=13.5cm]{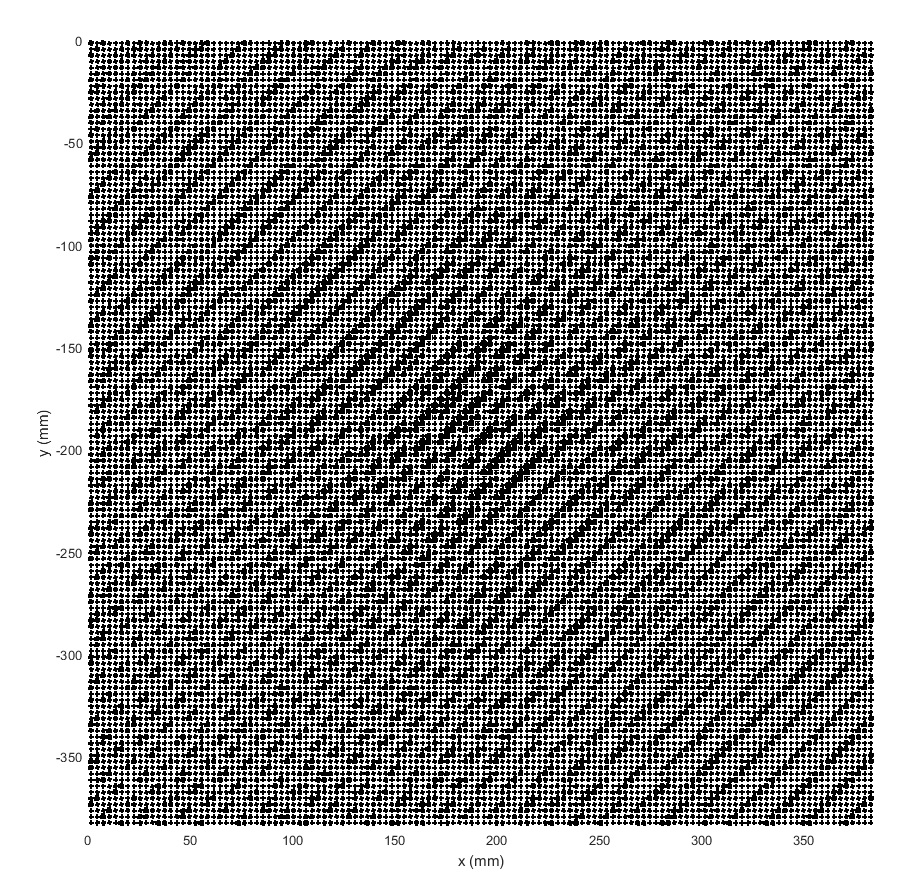}
\caption{Holographic metasurface of the CGH of Bessel beam implemented using unit cells with gaps variation.}
\label{f5}
\end{figure}

\h For this first HMS, we reproduce the CGH of a Bessel wave of zero order, with a resolution of 128x128 pixels, transverse number wave $k_{\rho}=16\ \textrm{mm}^{-1}$, size of central spot of 0.28 mm and generated at wavelength of $\lambda=12.33$ mm, corresponding to our operating frequency of 24.34 GHz (see Figure \ref{f4}), the corresponding holographic metasurface was built applying the theory presented above and it is shown in the Figure \ref{f5}. We can note the small unit cells with variations of gap, the darkest zones correspond to cells with smaller gaps or high surface impedance whilst the clearest zones correspond to cells with larger gaps or low surface impedance. \\

\begin{figure}[H]
\centering
\includegraphics[width=5cm]{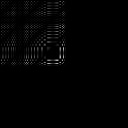}
\caption{Image of the computer-generated hologram of an Airy beam with resolution of 128x128 pixels at wavelength $\lambda=12.33$ mm.}
\label{f6}
\end{figure}

\begin{figure}[H]
\centering
\includegraphics[width=13.5cm]{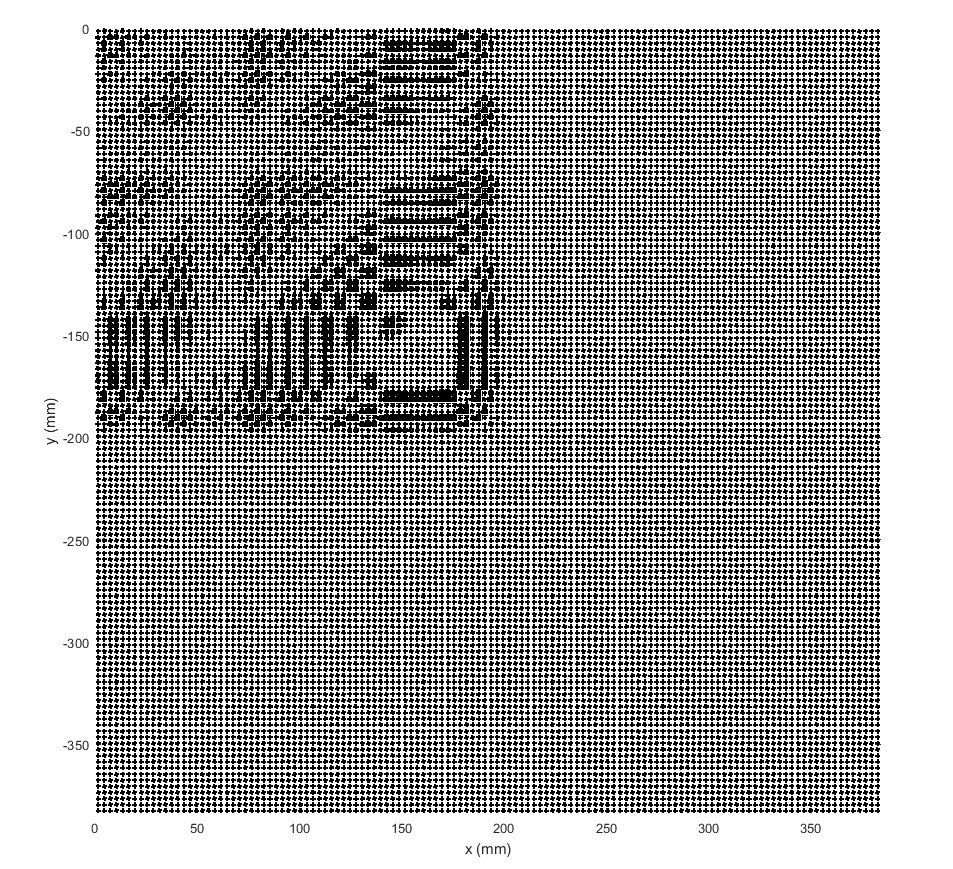}
\caption{Holographic metasurface of the CGH of an Airy beam implemented using unit cells with gaps variation.}
\label{f7}
\end{figure}

\begin{figure}[H]
\centering
\includegraphics[width=5.5cm]{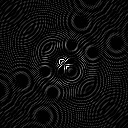}
\caption{Image of the computer-generated hologram of a Frozen Wave (FW) beam with resolution of 128x128 pixels at wavelength $\lambda=12.33$ mm.}
\label{f8}
\end{figure}

\begin{figure}[H]
\centering
\includegraphics[width=14cm]{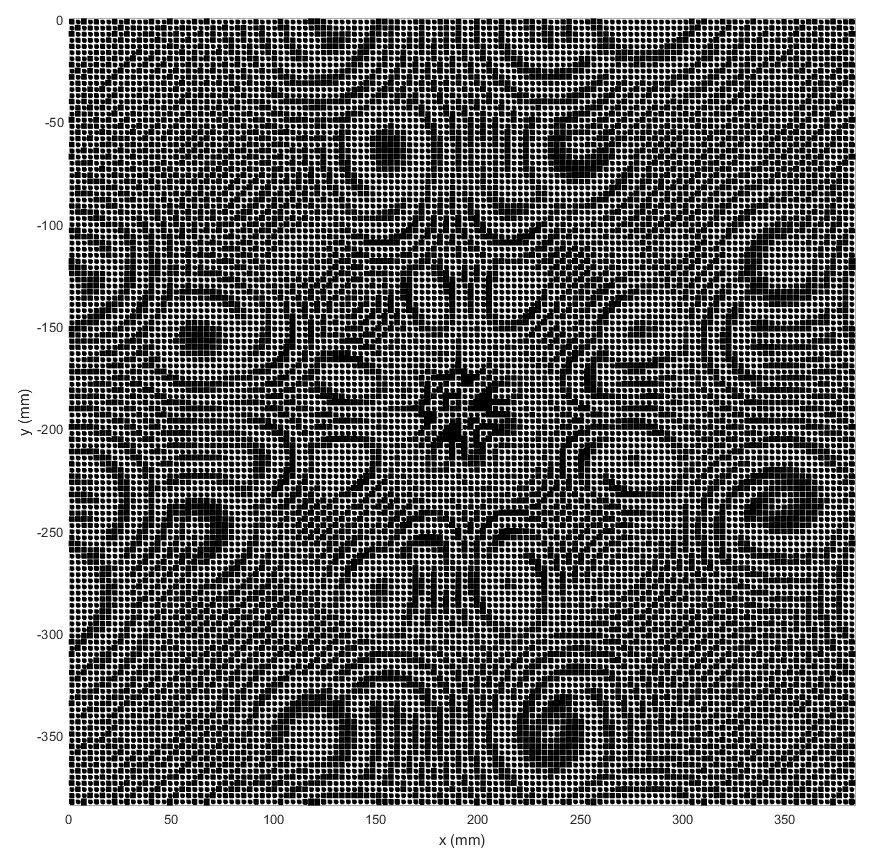}
\caption{Holographic metasurface of the CGH of FW beam implemented using unit cells with gaps variation.}
\label{f9}
\end{figure}

\h We also reproduce the CGH of an Airy wave, with a resolution of 128x128 pixels, parameter of decay $a=0.1$ \cite{ref:27} and generated at wavelength of $\lambda=12.33$ mm, corresponding to our operating frequency of 24.34 GHz (see Figure \ref{f6}), the corresponding HMS is shown in the Figure \ref{f7}. \\

\h And, we also reproduce the CGH of a Frozen Wave (FW), with a resolution of 128x128 pixels, number of Bessel beams superposed $N=6$, size of central spot $\Delta\rho_{0}=7.8$ mm, constant $Q=407.67$ \cite{ref:30} and generated at wavelength of $\lambda=12.33$ mm, corresponding to our operating frequency of 24.34 GHz (see Figure \ref{f8}), the corresponding HMS is shown in the Figure \ref{f9}. \\

\begin{figure}[H]
\centering
\includegraphics[width=13cm]{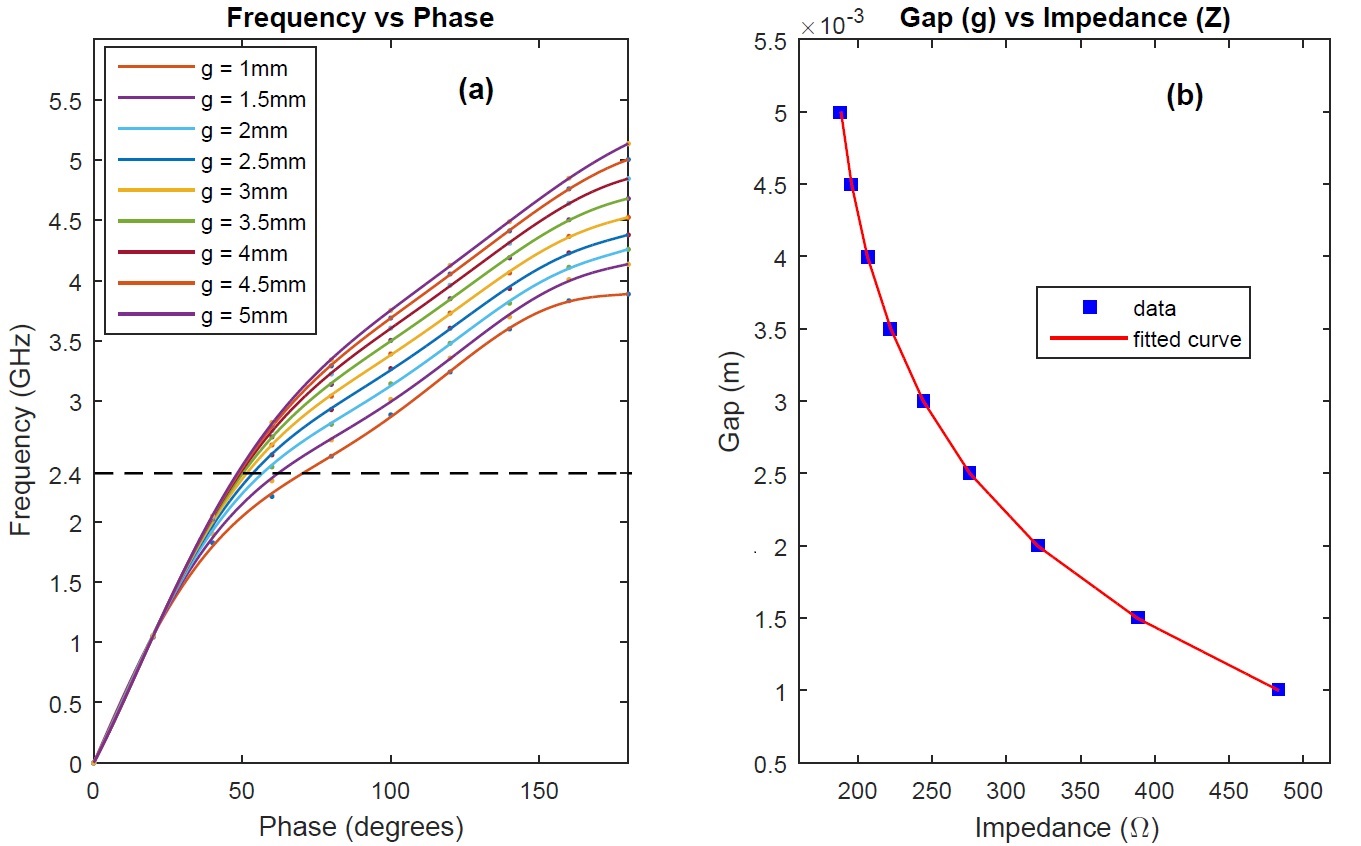}
\caption{(a) Variations of frequency with phase for each value of gap ($g$), the operating frequency was defined to 2.4 GHz. (b) Variation of values of gaps according to surface impedances and its corresponding adjustment curve.}
\label{f10}
\end{figure}

\h A second holographic metasurface was built by simulation using CST Microwave Studio. This case, the operating frequency is 2.4 GHz, the metal used was copper (Cu) on a substrate Rogers TM6 with permittivity $\epsilon=6$ and thickness $t=7.85$ mm. The lattice parameter was set at $d=15$ mm, and gaps were set from $g=1$ mm till $g=5$ mm with an interval of 0.5 mm. The range of frequencies in the simulation was defined from 0 till 5.5 GHz. In the four sides of the unit cell (see Figure \ref{f1}) were set periodic boundary conditions and Perfect Electrical Conductor (PEC) for the propagation direction. The graph of the dispersion curves for each value of gap and variation of gaps with surface impedance are shown in the Figure \ref{f10}. After making the adjustment of the curve of gaps ($g$) vs surface impedance ($Z$), we obtained the interval for surface impedance: $[Z_{\textrm{min}}, Z_{\textrm{max}}] = [188.6\ \Omega, 483.4\ \Omega]$, and we also found the following modulate values: $X=188.6\ \Omega$, $M=46$.

\begin{figure}[H]
\centering
\includegraphics[width=5.5cm]{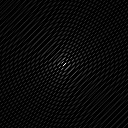}
\caption{Image of the computer-generated hologram of a Bessel beam with resolution of 128x128 pixels at wavelength 125 mm.}
\label{f11}
\end{figure}

\begin{figure}[H]
\centering
\includegraphics[width=14cm]{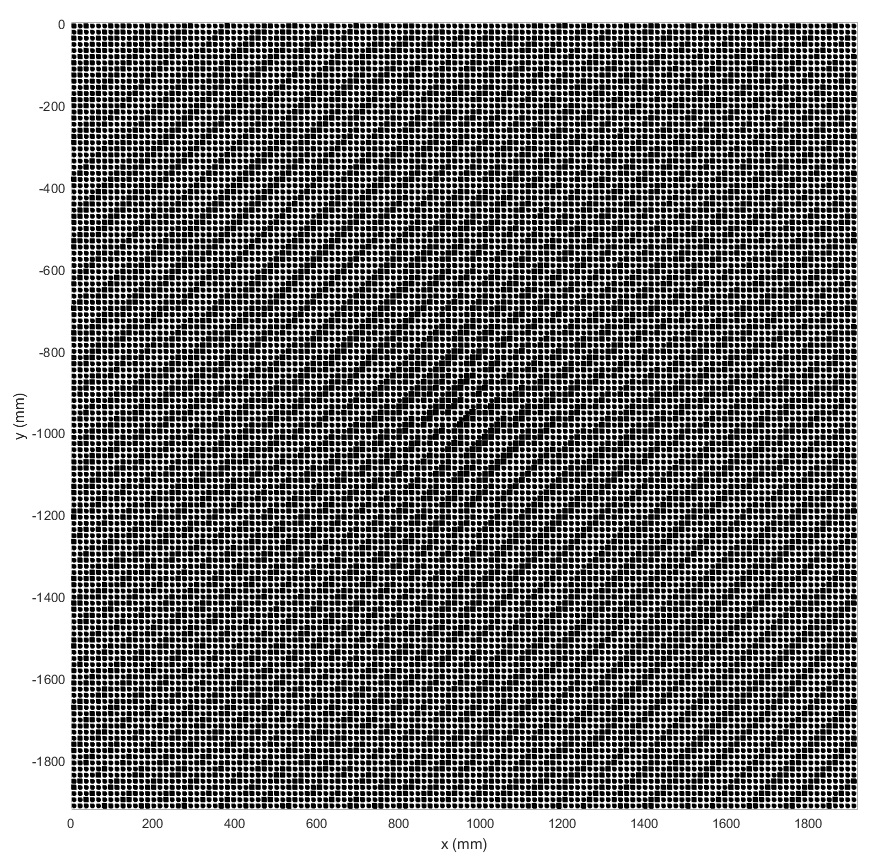}
\caption{Holographic metasurface of the CGH of Bessel beam implemented using unit cells with gaps variation.}
\label{f12}
\end{figure}

\begin{figure}[H]
\centering
\includegraphics[width=5.5cm]{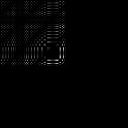}
\caption{Image of the computer-generated hologram of an Airy beam with resolution of 128x128 pixels at wavelength 125 mm.}
\label{f13}
\end{figure}

\begin{figure}[H]
\centering
\includegraphics[width=14cm]{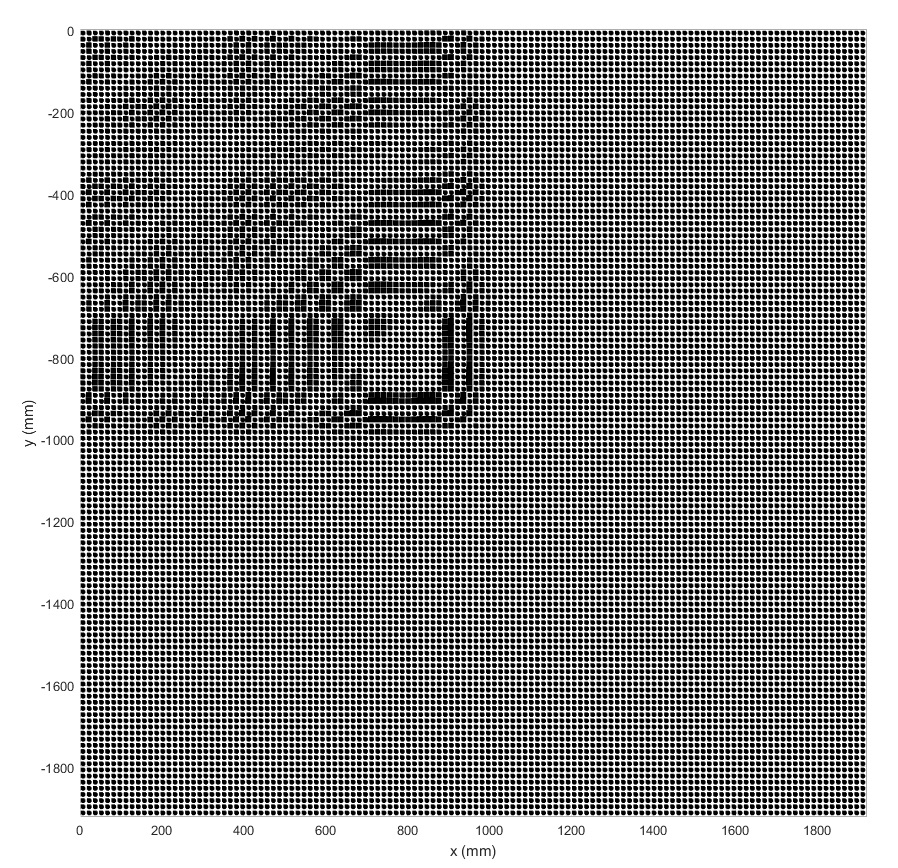}
\caption{Holographic metasurface of the CGH of Airy beam implemented using unit cells with gaps variation.}
\label{f14}
\end{figure}

\begin{figure}[H]
\centering
\includegraphics[width=5.5cm]{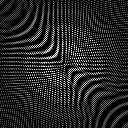}
\caption{Image of the computer-generated hologram of a frozen wave beam with resolution of 128x128 pixels at wavelength 125 mm.}
\label{f15}
\end{figure}

\begin{figure}[H]
\centering
\includegraphics[width=14cm]{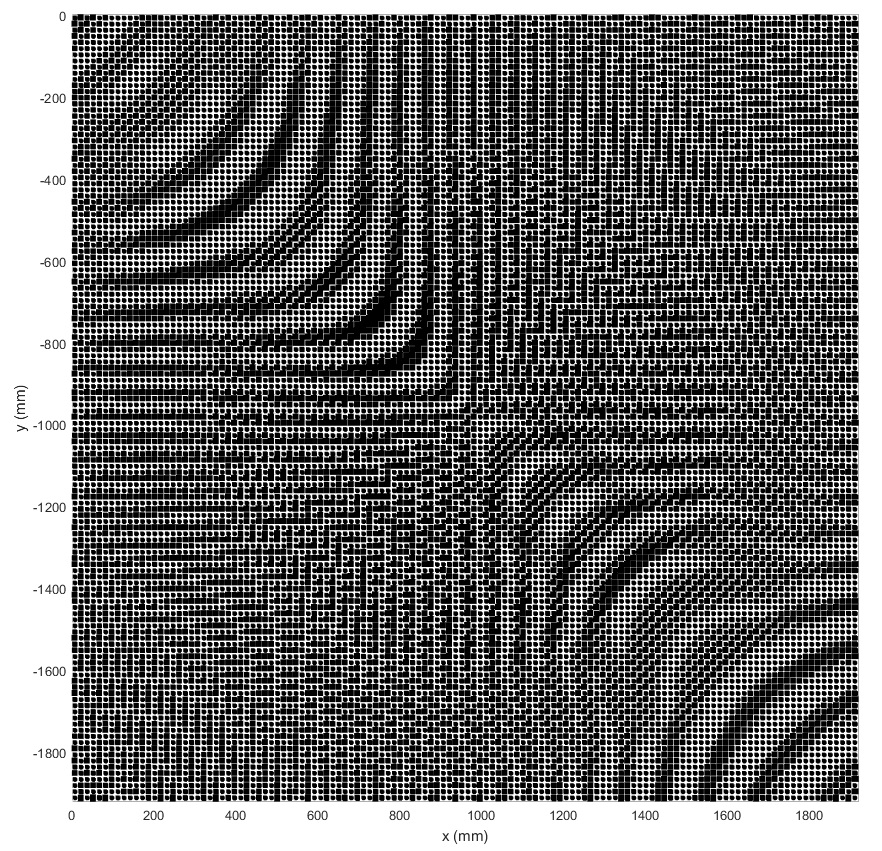}
\caption{Holographic metasurface of the CGH of FW beam implemented using unit cells with gaps variation.}
\label{f16}
\end{figure}

\h For this second HMS, we reproduce the CGH of a Bessel wave of zero order, with a resolution of 128x128 pixels, transverse number wave $k_{\rho}=16\ \textrm{mm}^{-1}$, size of central spot of 0.28 mm and generated at wavelength of $\lambda=125$ mm, corresponding to our operating frequency of 2.4 GHz (see Figure \ref{f11}), the corresponding holographic metasurface is shown in the Figure \ref{f12}.\\

\h Figure \ref{f13} shows the CGH of an Airy wave, with a resolution of 128x128 pixels, parameter of decay $a=0.1$ \cite{ref:27}, the corresponding HMS is shown in the Figure \ref{f14}.\\

\h Finally, we reproduce the CGH of a Frozen Wave (FW), with a resolution of 128x128 pixels, number of Bessel beams superposed $N=6$, size of central spot $\Delta\rho_{0}=47.9$ mm, constant $Q=4.52$ \cite{ref:30}, in Figure \ref{f15}), and the corresponding HMS is shown in the Figure \ref{f16}.

\h {\em 4. Conclusions} --- This work presents a way for controlling and manipulating the electromagnetic radiation through the computational realization of holographic metasurfaces to generation of the non-diffracting waves. Holographic metasurfaces (HMS) are simulated by modeling a periodic lattice of metallic patches on dielectric substrates with sub-wavelength dimensions, where each one of those unit cells alter the phase of the incoming wave. The surface impedance (Z) allows to control the phase of a wave through the metasurface in each unit cell. The sub-wavelength dimensions guarantees that the effective medium theory is fulfilled. The metasurfaces are designed by the computer-generated hologram (CGH) of non-diffracting waves are generated and reproduced using such HMS in the microwave regime. Two sets of holographic metasurfaces was built for working in the microwave regime, the first at 24.34 GHz and the second one at 2.4 GHz. The results is according to the theoretically predicted and allows applications of these types of electromagnetic waves in several areas of telecommunications and bioengineering. \\

\h {\em Acknowledgments} The authors acknowledge partial support from UFABC, CAPES, FAPESP (UNDER GRANTS 16/19131-6) and CNPq (UNDER GRANTS 302070/2017-6).


\begin{thebibliography}{99}

\bibitem{ref:1} J. Joannopoulos, S. Johnson, J. Winn, R. Meade. ``Photonic Crystals. Molding the flow of light''. Princeton University Press, second edition, 2008.

\bibitem{ref:2} E. Yablonovitch. ``Photonic band-gap crystals''. Journal of Physics Condensed Matter \textbf{5}, 2443 (1993).

\bibitem{ref:3} V. Veselago. ``The Electrodynamics of Substances with simultaneously negative values of $\varepsilon$ and $\mu$''. Soviet Physics Uspekhi \textbf{10}, 4 (1968).

\bibitem{ref:4} Y. Liu, X. Zhang. ``Metamaterials: a new frontier of science and technology''. Chemical Society Reviews \textbf{40}, 2494-2507 (2011).

\bibitem{ref:5} S. Ramakrishna. ``Physics of negative refractive index materials''. Reports on Progress in Physics \textbf{68}, 449-521 (2005).

\bibitem{ref:6} L. Billings. ``Metamaterial World''. Nature \textbf{500}, 138 (2013).

\bibitem{ref:6a} D. Smith \textit{et al}. ``Composite Medium with Simultaneously Negative Permeability and Permittivity''. Physics Review Letters. \textbf{84}, 4184 (2000)

\bibitem{ref:7} D. Smith, N. Kroll. ``Negative Refractive Index in Left-Handed Materials''. Physical Review Letters. \text{85}, 2933 (2000).

\bibitem{ref:7a} J. Pendry. ``Negative Refraction Makes a Perfect Lens''. Physical Reviews Letters \textbf{85}, 3966 (2000).

\bibitem{ref:7b} X. Zhang, Z. Liu. ``Superlenses to overcome the diffration limit''. Nature materials \textbf{7}, 435 (2008).

\bibitem{ref:8} D. Smith \textit{et al}. ``Metamaterials and Negative Refractive Index''. Science \textbf{305}, 788 (2004)

\bibitem{ref:9} R. Shelby \textit{et al}. ``Experimental Verification of a Negative Index of Refraction.'' Science \textbf{292}, 77 (2001)

\bibitem{ref:10} H. Lezec, J. Dionne, H. Atwater. ``Negative Refraction at Visible Frequencies.'' Science \textbf{316}, 430 (2007)

\bibitem{ref:11} J. Pendry \textit{et al}. ``Controlling Electromagnetic Fields''. Science \textbf{312}, 1780 (2006).

\bibitem{ref:12} D. Schuring \textit{et al}. ``Metamaterial Electromagnetic Cloak at Microwave Frequencies''. Science \textbf{314}, 977 (2006)

\bibitem{ref:13} C. Holloway \textit{et al}. ``A discussion on the interpretation and characterization of metafilms/metasurfaces: The two-dimensional equivalent of metamaterials''. Metamaterials \textbf{3}, 100-112 (2009).

\bibitem{ref:14} H. Chen, A. Taylor, N. Yu. ``A review of metasurfaces: physics and applications''. Reports on Progress in Physics \textbf{79}, 076401 (2016)

\bibitem{ref:15} N. Yu \textit{et al}. ``Light propagation with phase discontinuities: generalized laws of reflection and refraction''. Science \textbf{334}, 333 (2011)

\bibitem{ref:16} F. Aieta \textit{et al}. ``Reflection and refraction of light from metasurfaces with phase discontinuities''. Journal of Nanophotonics \textbf{6}, 063532 (2012)

\bibitem{ref:17} A. Kildishev, A. Boltasseva, V. Shalaev. ``Planar Photonics with Metasurfaces''. Science \textbf{339}, 1232009 (2013).

\bibitem{ref:18} L. Liu \textit{et al}. ``Broadband Metasurfaces with Simultaneous Control of Phase and Amplitude''. Advanced Materials Banner \textbf{26}, 5031 (2014).

\bibitem{ref:19} M. Kats \textit{et al}. ``Giant birefringence in optical antenna arrays with widely tailorable optical anisotropy''. Proceedings of the National Academy of Sciences of USA \textbf{109}, 12364 (2012).

\bibitem{ref:20} F. Zhou, Y. Liu, W. Cai. ``Plasmonic holographic imaging with V-shaped nanoantenna array''. Optics Express \textbf{21}, 4348 (2013).

\bibitem{ref:21} B. Fong \textit{et al}. ``Scalar and Tensor Holographic Artificial Impedance Surfaces''. IEEE Transactions on Antennas and Propagation \textbf{58}, 3212 (2010).

\bibitem{ref:22} Y. Li, X. Wan, B. Cai \textit{et al}. ``Frequency-Controls of Electromagnetic Multi-Beam Scanning by Metasurfaces''. Scientific Reports \textbf{4}, 6921 (2014).

\bibitem{ref:23} B. Saleh, M. Teich. ``Fundamentals of Photonics''. John Wiley \& Sons, INC (1991)

\bibitem{ref:24} J. Goodman. ``Introduction to Fourier Optics''. Roberts \& Co. Publishers, 2 ed. (2004)

\bibitem{ref:25} P. Hariharan. ``Optical Holography''. Cambridge University Press, 2 ed. (1996)

\bibitem{ref:26} G. Siviloglou and D. Christodoulides. ``Accelerating finite energy Airy beams''. Optics Letters \textbf{32}, 979 (2007).

\bibitem{ref:28} N. Chattrapiban \textit{et al}. ``Generation of nondiffracting Bessel beams by use of a spatial light modulator''. Optics Letters \textbf{28}, 2183 (2003).

\bibitem{ref:29} J. Durnin, J. Miceli and J. Eberly. ``Comparison of Bessel and Gaussian beams''. Optics Letters \textbf{13}, 79 (1988).

\bibitem{ref:30} M. Zamboni-Rached. ``Stationary optical wave fields with arbitrary longitudinal shape by superposing equal frequency Bessel beams: Frozen Waves''. Optics Express \textbf{12}, 4002 (2004).

\bibitem{ref:31} T. A. Vieira, M. R. R. Gesualdi, M. Zamboni-Rached. ``Frozen waves: experimental generation''. Optics Letters \textbf{37}, 2034 (2012).

\bibitem{ref:32} T. A. Vieira, M. Zamboni-Rached, M. R. R. Gesualdi. ``Modeling the spatial shape of nondiffracting beams: Experimental generation of Frozen Waves via holographic method''. Optics Communications \textbf{315}, 374 (2014).

\bibitem{ref:33} T. A. Vieira, M. R. R. Gesualdi, M. Zamboni-Rached and E. Recami. ``Production of \textit{dynamic} frozen waves: controlling shape, location (and speed) of diffraction-resistant beams''. Optics Letters \textbf{40}, 5834 (2015).

\bibitem{ref:34} E. G. P. Pachon, M. Zamboni-Rached, A. Dorrah, M. Mojahedi, M. R. R. Gesualdi and G. G. Cabrera. ``Architecting new diffraction-resistant light structures and their possible applications in atom guidance''. Optics Express \textbf{24}, 25403 (2016).

\bibitem{ref:35} R. A. B. Suarez, T. A. Vieira, I. S. V. Yepes, M. R. R. Gesualdi. ``Photorefractive and computational holography in the experimental generation of Airy beams''. Optics Communications \textbf{366}, 291 (2016).

\bibitem{ref:36} T. A. Vieira, R. A. B. Suarez, I. S. V. Yepes, M. R. R. Gesualdi, M. Zamboni-Rached. ``Optical reconstruction of non-diffracting beams via photorefractive holography''. Applied Physics. B, Lasers and Optics \textbf{123}, 134 (2017).

\bibitem{ref:37} M. Zamboni-Rached, E. Recami, T. A. Vieira, M. R. R. Gesualdi, J. Nobre-Pereira. ``Structured Light by Linking Diffraction-Resistant Spatially Shaped Beams''. Physical Review Applied \textbf{10}, 034023 (2018).

\bibitem{ref:39} T. A. Vieira, M. Zamboni-Rached, M. R. R. Gesualdi. ``Experimental Generation of Frozen Waves in Optics: Control of Longitudinal and Transverse Shape of Optical Non-diffracting Waves. In: H. E. Hernández-Figueroa, E. Recami and M. Zamboni-Rached. (Org.). Non-Diffracting Waves``. Wiley-VCH Verlag, 417 (2013).

\bibitem{ref:40} S. R. C. Fernandez, M. R. R. Gesualdi. ``Holographic metasurfaces applied to generation of non-diffracting beams''. Latin America Optics and Photonics Conference. Washington: OSA \textbf{Th4A.22} (2018).


\end{thebibliography}
\end{document}